\documentstyle[prl,multicol,aps]{revtex}
\input{epsf}
\begin{document}
\draft
\title{The Fano Effect in Aharonov-Bohm interferometers
\thanks{Dedicated to Peter W\"olfle on the occasion of his 60th
birthday} }
\author{O. Entin-Wohlman$^a$, A. Aharony$^a$, Y. Imry$^b$ and Y.
Levinson$^b$}
\address{$^a$School of Physics and Astronomy, Raymond and Beverly Sackler
Faculty of Exact Sciences, \\ Tel Aviv University, Tel Aviv 69978,
Israel\\ }
\address{$^b$Department of Condensed Matter Physics, The Weizmann
Institute of Science, Rehovot 76100, Israel}

\date{\today}
\maketitle
\begin{abstract}

After briefly reviewing the Fano effect, we
explain why it may be relevant to various types of Aharonov-Bohm
interferometers.
We discuss both
closed (electron conserving) and open
interferometers, in which one path contains either a simple quantum dot or
a decorated quantum dot (with more than one internal state or a parallel path).
The possible relevance  to some
hitherto unexplained experimental features is also discussed.

\end{abstract}



\section{Introduction}

In this paper we consider solid-state interferometers, restricted
to the mesoscopic scale in order to retain the coherence of the
conduction electrons\cite{Book}. Here the two-slit geometry is
often replaced by narrow waveguides, possibly containing
scatterers \cite{Land}, for the electron paths. An Aharonov-Bohm
(AB) \cite{AB} flux $\Phi$ between the two paths in such interferometers \cite{GIA,Webb}
yields a conductance $G(\Phi)$ which contains an interference
term proportional to $\cos(\phi+\beta)$,
where $\phi = 2\pi \Phi/\Phi_0$, and $\Phi_0=hc/e$ is the single-electron
flux quantum. Experiments with a quantum dot (QD) on one of
the interfering paths aim to relate $\beta$ to the dot's intrinsic
\cite{Ng} Friedel \cite{Langreth} transmission phase, $\alpha_1$.
For closed systems, which conserve the electron current
(unitarity), time-reversal symmetry  requires via the Onsager
relation \cite{onsager} that $G(\Phi)=G(-\Phi)$, and therefore that
$\beta=0$ or $\pi$. Thus, to measure a non-trivial value of $\beta$ one has to open
up the system in order to break unitarity.  For open systems, it
was recently shown \cite{Entin-Wohlman} that $\beta$ depends in
general on the details of the broken unitarity. Specific ways of
opening the system were discussed, so that
the transmission
amplitude through the two paths is equal to the sum of the transmission amplitudes
through the two individual paths, as in the textbook two-slit geometry
\cite{Feynmann,Schwabl,Amir1}. In this case, $\beta$ is equal to the phase difference
between these two amplitudes, $\alpha_1-\alpha_2$, and one gets direct information on
$\alpha_1$.

The AB $h/e$ oscillations in $G(\Phi)$, first suggested (in spite of
strong impurity scattering) in Ref. \onlinecite{GIA}, were
subsequently observed on metallic closed systems\cite{Webb} and in
semiconducting samples containing QDs near Coulomb blockade (CB)
resonances \cite{Amir2,Yacoby3}. In these experiments
$G(\Phi)=G(-\Phi)$, as required by the Onsager symmetry. Further
experiments \cite{Yacoby3,Eyal,Schuster,Ji1,Ji2} used {\it open
systems}, in which electrons are lost (going to other electron reservoirs)
from the transmitted current, to obtain a non-zero phase
shift $\beta$. Assuming that $\beta=\alpha_2-\alpha_1$, some of
the surprising experimental results were inconsistent with the
theoretical expectations for the phase $\alpha_1$ of the intrinsic
transmission through the QD\cite{Ng,Langreth,Oreg}. Examples
include period doubling for the oscillation as function of the
gate voltage \cite{Amir2}, the phase lapse between consecutive CB
resonances \cite{Eyal,Schuster} and the non-universal phase shifts
\cite{Langreth,Oreg} at the Kondo resonances\cite{Ji1,Ji2}. These
findings have, as yet, not received a universally accepted
explanation.

As discussed in Ref. \cite{Entin-Wohlman}, even the unitary
(current conserving) interferometer (shown in Fig. 1a) reveals
very useful information on the resonances of the QD: although the
dependence of $G$ on $\phi$ is only via $\cos\phi$, so that
$\beta=0$ or $\beta=\pi$, the coefficient of $\cos\phi$ in $G$
changes sign near each resonance, and this could be described as
``jumps" of $\beta$ from zero to $\pi$ or vice versa. In addition
to the vanishing of this coefficient, it turns out that the whole
conductance may also vanish (or become very small) at special
values of the parameters. For an appropriate choice of parameters,
when one may use a perturbative calculation, this also happens
close to the resonances. As we discuss below, these zeroes are
directly related to a destructive interference between the two
paths, in close analogy with the Fano effect
\cite{Fano,Hist,Bulka,Hofs}, which concerns destructive
interference in the absorption lines between discrete levels and
 the continuum. In the present paper we elaborate on these
phenomena.

Having discussed the elementary closed interferometer, we then consider a more
complex situation.
We introduce
a miniature version of the previous closed ring (made of two parallel paths, with a QD
on one or on both), as our effective scattering element. We call this combined element
the decorated quantum dot (DEQD). We place this DEQD on one path of a larger
AB interferometer, as shown in Fig. 1b.
(Another example of such a DEQD is a QD with two resonances.)
The DEQD, which now replaces the simpler QD of Fig. 1a,
captures situations in which one can have internal
interference inside the QD.
The Fano zeroes   of the DEQD have a profound effect on
the  interference
part of the conductance of the whole AB interferometer consisting
of the DEQD and the other conducting branch of the AB
interferometer.
 Opening the larger interferometer, so that it obeys the two-slit
rules, this procedure allows a measurement of the phase shift
associated with the DEQD. The latter is by itself  sensitive to
the Fano effect. This ``hierarchical Fano effect" is the main
subject of the present paper. Note that the area subtended between
the conducting branch and the DEQD branch is finite and the flux
through it is significant, unlike that of the ``miniature"
conducting branch of the DEQD itself.


In Sec. {\ref{Fano} we briefly review the original Fano derivation \cite{Fano}, and
present an argument for its physical relevance for AB
interferometers. We then present calculations on two types of  AB interferometers
which exhibit the Fano effect and demonstrate the above  general arguments.
In Sec. III we follow Ref.
\onlinecite{GIA}, treating each scattering element via its scattering matrix.
A simple tight binding model, based on Ref. \onlinecite{Entin-Wohlman}, is
presented in section \ref{Model} (see fig. 1c). Some technical details of
the solution are given
in the appendix
The consequences of such models are argued in
section \ref{Physics} to be rather generic, and results from various
other models are cited to support that. The possible relevance
to experiments on AB interferometers is then briefly
discussed.

\section{Fano Physics
} \label{Fano}

In 1961, Fano \cite{Fano} considered a physical situation which
occurs in many  systems addressed spectroscopically. In the first
part of the calculation, a quantum state, $|\phi\rangle$, with
energy $E_\phi^0$, is taken to be coupled via matrix elements
$V_{E'}$ to a continuum of states denoted by $|E'\rangle$, forming
a resonance in the continuum. In the second part, a transition
between another state, $|i \rangle$, (which is not resonant with
the continuum) and the resonance is considered, and it is shown that the interference
between the contributions of the original continuum states and the resonance will always
yield an energy at which the total transition amplitude vanishes. Fano also treated
various generalizations which will not be discussed here.
In this section we first give a brief review of Fano's derivation,
and then discuss its relevance to interferometers.

\subsection{Fano's derivation}

We start with a quasicontinuum having $N >> 1$ nondegenerate
states, with a very small, roughly uniform, level separation $d$
and a corresponding density of states (DOS) equal to $1/d$.
The exact eigenstate with energy ${\cal E}$ is written as
\begin{equation}
|\Psi_{\cal E} \rangle = a_{\cal E} |\phi\rangle + \sum  b_{{\cal E},E'} |E'\rangle.
\label{eigens}
\end{equation}
The Schr\"odinger equation for this state becomes
\begin{eqnarray}
E_\phi^0 a_{\cal E} + \sum  b_{{\cal E},E'} V_{E'}^*= {\cal E}a_{\cal E}, & \nonumber \\
a_{\cal E} V_{E'} + b_{{\cal E},E'} E'= {\cal E} b_{{\cal E},E'}. \label{Schr}
\end{eqnarray}

For a discrete spectrum of $E'$, no matter how dense, the well
known solution of these equations is found by
solving the second equation, $b_{{\cal E},E'}=a_{\cal E}V_{E'}/({\cal E}-E')$,
and plugging  this
value into the
first one.
The new energies ${\cal E}$ are then obtained from the equation
\begin{equation}
{\cal E} = E_\phi^0 +   \sum |V_{E'}|^2 / ({\cal E} - E').
\label{newE}
\end{equation}
It has already been found by Rayleigh that $N-1$ of the new
energies, ${\cal E}$, straddle the old ones $E'$, and two states are
outside of the original band. Assuming that no localized states
occur at finite distances from the band edges, there is a
resonance within the band, centered near ${\cal E}_\phi$ (see below).
 This resonance alters the total DOS
everywhere
only by relative order $1/N)$,
and the $b_{{\cal E},E'}$'s have significant values over a width
$({\cal E-E}_\phi)$ of order $V_{{\cal E}_\phi}$.
The last term in Eq. (\ref{newE}) represents the ``self consistent" repulsion of the
exact energy ${\cal E}$ by the band states. For a symmetric $|V_E|$ and a symmetric
band, this term vanishes in the middle of the band.
A similar level repulsion shifts the peak of the resonance from $E_\phi^0$ to
${\cal E}_\phi$.

We now go to the continuum limit, where $d \rightarrow 0$, $N
\rightarrow \infty$ and $Nd$ remains finite, yielding a $O(N)$
density of states $\rho(E) \sim \frac{1}{d}$. In this limit one also has
$V_E \rightarrow 0$ but
the combination
\begin{equation}
\gamma(E) =  |V_{E}|^2 \rho(E)
\end{equation}
remains finite. The sums now go into integrals with the DOS,
which will basically convert $|V_{E'}|^2$ in the sums into
$\gamma$ in the integrals. The nontrivial mathematical point in
solving Eq. (\ref{Schr}) in the continuum limit, is in the
inversion of ${\cal E} - E'$. This is what Fano did extremely
carefully. Formally, the solution for $b_{{\cal E},E'}$ from the
second equation, can be written as
\begin{equation}
b_{{\cal E},E'} = [{\cal P} \frac{1}{{\cal E} - E'} + z({\cal E}) \delta ({\cal E} - E')
]V_{E'}a_{\cal E}.
\label{bEE}
\end{equation}
One is used to writing $z({\cal E}) = i \pi$. However, as pointed
out by Fano, for stationary states, one must {\it determine}
$z({\cal E})$ from Eqs. (\ref{Schr}). The first of these
equations, corresponding to Eq. (\ref{newE}) in the discrete case,
becomes then
\begin{equation}
{\cal E} = E_\phi^0 + z({\cal E}) \gamma + {\cal P} \int dE' \gamma (E')/ ({\cal E} - E')
\end{equation}
and $z({\cal E})$ is given by
\begin{equation}
z({\cal E}) = \frac{{\cal E}- E_\phi^0-{\cal P}
\int dE'\gamma (E')/({\cal E}-E')}{\gamma (E)}.
\label{z}
\end{equation}
The principal part of the integral is analogous to the sum in the
discrete case and gives the shift \cite{Fano} of the resonance, from $E_\phi^0$ to the new
value ${\cal E}_\phi$, where $z({\cal E}_\phi)=0$. The fact that
$z$ is real is very significant. Only systems satisfying
time-reversal symmetry, in which the Hamiltonian matrix is real and
symmetric, are considered, and ${\cal E}$ is strictly on the real axis.
Moreover, an explicit calculation of the normalization factor, $|a_{\cal E}|^2$ \cite{Fano},
 gives a Lorentzian which is centered at ${\cal E}_\phi$, with width
$\pi\gamma$ (we take $\gamma$ to be much smaller than the
bandwidth, which is usually of the order of $Nd$).
Thus, $z({\cal E})$ changes sign at
the actual center of the resonance. These facts provide the
mathematical basis for the nontrivial effects found by Fano.

As an example of these effects, Fano considers the absorption from
some given initial (non-resonating) state, $|i\rangle$, into the
state $|\Psi_{\cal E} \rangle$. Naively, one expects the usual
Lorentzian line shape in the absorption, due to the local DOS,
which in turn is the total DOS, weighed for each energy by
$|a_{\cal E}|^2$. This does in fact happen for a transition
operator \footnote{It might seem to the casual reader that  $\hat
T$ is just the portion of the Hamiltonian, say ${\cal H}_1$, that
causes the transition. In this case the transition rate calculated
would have been just the one to lowest order in ${\cal H}_1$.
However, the understanding that $\hat T$ is the full transition
opereator implies that all higher order processes have been summed
upon, which converts ${\cal H}_1$ into $\hat T$. In the closed
interferometer the difference between using just ${\cal H}_1$ and the
full  $\hat T$ amounts to going from the naive two-path
expression for the transmission amplitude [for example, the last
brackets of the third equality in Eq. (\ref{amplitudes})] to its
full expression there.} $\hat T$, that has matrix elements
$\langle\phi| \hat T |i \rangle$  {\it only} with the state $|\phi
\rangle$. Once matrix elements with the original continuum states
also exist, $\langle E'| \hat T |i \rangle \ne 0$, interference
between the two transition amplitudes from $|i\rangle$ to $|\phi
\rangle$ and to $|E' \rangle$ will occur. Substituting Eq.
(\ref{bEE}) into Eq. (\ref{eigens}) and turning the sum into an
integral, one has
\begin{equation}
\langle i|\hat T|\Psi_{\cal E} \rangle=a_{\cal E}(\langle i|\hat T|\psi \rangle +
\rho({\cal E})V_{\cal E} z({\cal E}) \langle i| \hat T|{\cal E} \rangle),
\label{iTPsi}
\end{equation}
where
\begin{equation}
|\psi \rangle  = |\phi \rangle + {\cal P} \int dE'
\rho(E')\frac{V_{E'}
 |E' \rangle}{{\cal E} - E'}
\end{equation}
indicates the original state $|\phi \rangle$ modified by the coupling to the continuum.
It is important to note the effect of adding the second term in Eq. (\ref{iTPsi}): Since $z({\cal E})$ is real, {\it
with opposite signs on the two sides of the resonance}, and since
(for a system satisfying time-reversal symmetry) the matrix elements can also be chosen
to be real, {\it the net transition amplitude will always vanish} at some energy ${\cal E}_0$
which satisfies
\begin{equation}
 z({\cal E}_0) = -\frac{\langle i| \hat T |\psi \rangle}{\rho({\cal E}_0)
V_{{\cal E}_0}\langle i|
 \hat T |{\cal E}_0 \rangle}.
\end{equation}
This will not
be the case for a complex expression (such as that obtained when a
magnetic field is applied) as a function of a single real
parameter. It has zeroes on the real axis with zero probability.
Eq. (\ref{iTPsi}) also
modifies the Lorentzian absorption lineshape. The opposite relative
phases of the two terms on the two sides of the resonance
turn the Fano absorption lineshape asymmetric around the resonance.
Such a Fano lineshape occurs very frequently in the spectroscopy of various
atomic, molecular and solid state systems.

\subsection{Fano effect in the closed interferometer}

\begin{figure}
\leavevmode \epsfclipon \epsfxsize=8truecm
\vbox{\epsfbox{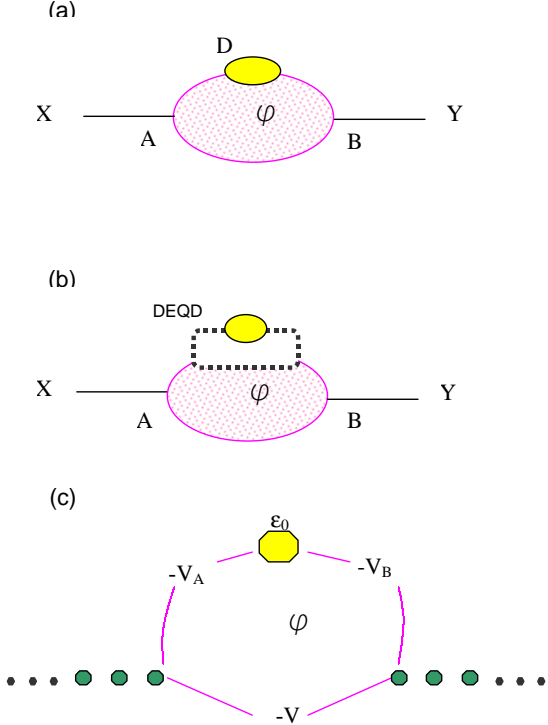}} \caption{{\bf (a)} A generic model for a
closed AB interferometer having a QD denoted by $D$ on one of its
arms. The whole closed interferometer is our DEQD.
{\bf (b)} The DEQD in parallel with the conducting arm of the AB
interferometer. {\bf (c)} The tight-binding model for the DEQD.}
 \label{Int}
\end{figure}

We now  give a qualitative explanation why the transition, say,
from left to right across the interferometer  falls within the
Fano discussion. Consider the configuration in Fig. 1a, with a quantum dot D
which has one resonance state.
Clearly, such transition
rates will be governed by the full transition matrix element
between the last site on the left (A in the figure) to the first
one on the right (B in the figure). As in Sec. \ref{Model}
below, we denote the transfer matrix elements between A and D, D
and B and A and B respectively by $-V_A e^{i \phi}, -V_B$ and $-V$
(see Fig. 1c) . We take $V, V_A, V_B > 0$, $\phi$ is related to
the AB flux, as before, and a gauge putting all the vector
potential on the bond of $-V_A$ has been used here. Obviously the
results are gauge-invariant. Imagine, just for simplicity, that
$|V|, |V_A| << |V_B|$, so that the dot D is mainly coupled to the
right lead and all the couplings of the left lead are weak
perturbations. We first solve exactly the problem of the dot
coupled to the right lead, getting the usual resonance state of
the dot. Now, we calculate the transition rate from a given state
$|i\rangle$ on the left lead to the exact states of the dot--right
lead system. Each of the latter is an exact linear combination, as
in Eq. (\ref{eigens}), of the dot and the right-lead states. The
matrix elements between A and the former is $-V_A e^{i \phi}$, and with the
latter, it is $-V$. These two correspond  to $\langle\phi| \hat T
|i \rangle$ and $\langle E'| \hat T |i \rangle$
respectively. This establishes a full one-to-one correspondence
between the Fano problem and the AB inerferometer under
discussion. As explained above, Fano requires {\it real} transition amplitudes.
Therefore, the Fano effect will only be observed for $\phi=0$ or $\phi=\pi$.
In these cases,
the transition probability at some energy (which is proportional
to the transmission coefficient at the same energy), {\it must}
vanish on one side of the resonance.  ${\cal E}_0 - {\cal E}_{\phi}$
is found to have opposite signs at $\phi=0$ and $\phi=\pi$.

The above argument appears to have been  based on the lowest order golden rule, using only
the square of the transition matrix element. As we show below, the same effect appears
in the exact solution of this problem, where one sums over all the reflections
within the ring \cite{Entin-Wohlman,GIA}. This amounts to using, as in the original
Fano paper, the full transition operator $\hat T$ instead of the perturbation
Hamiltonian ${\cal H}_1$.
These exact results show that
increasing $|V|$ and $|V_A|$ can only
move the zero continuously as function of energy, but a zero will always exist
(unless interfered by other bands).

\subsection{Decorated QD in an AB interferometer}

Consider next the geometry in Fig. 1b. Here we replace the QD by a small ring,
which is itself similar to the whole ring of Fig. 1a. We call this small ring a ``decorated
QD" (DEQD). The DEQD is now placed on one path of the AB interferometer. The additional
path within the DEQD is aimed to allow internal interference, which usually exists
due to competing ``paths" through the QD. Assuming that the area of the
DEQD ring is much smaller than that of the full interferometer, we can neglect the flux
through the DEQD. The above Fano argument then implies that the
transmission amplitude of the DEQD branch of the AB interferometer, $t_{DEQD}$,
must vanish as function of energy in the vicinity of each resonance.

We now assume that the interferometer is
{\it open}, and that it obeys the necessary conditions for the two-slit equation to hold
\cite{Entin-Wohlman},
\begin{equation}
 T_{tot} = |t_{DEQD}e^{i \phi} + t_{par}|^{2},
\end{equation}
where $t_{par}$ is the transmission amplitudes of
the large parallel branch of the whole interferometer
(see Fig. 1b). We see that the two-path interference part of the AB
conductance,
\begin{equation}
 T_{int} = 2 {\cal R}(t_{DEQD}t_{par}^{*}e^{i\phi}),
\end{equation}
vanishes and changes sign at the energies of the DEQD Fano zeroes. Thus
the transmission phase of the DEQD, as measured  by the phase of
the AB oscillations, can be said to jump by $\pi$ at these points
\cite{Hist}.

Finally, we reemphasize the importance of going in the  golden-rule
type
calculation for the closed DEQD, from  a sum of lowest-order transition
amplitudes to one including all the multiple reflections \cite{GIA}.
The universal validity
of the Fano zeroes is  guaranteed as long as time-reversal
symmetry assures that all quantities in the calculation can  be
chosen real.
Interestingly, in this case, gauge invariance assures that  one
{\it may } use other choices. This is why deliberately not
choosing real parameters was not crucial in the calculation
reported below.

\section{Results on a simple closed interferometer}
\label{GIA}
\subsection{The DEQD}
In this chapter we consider the model of Ref. \cite{GIA}. The
transmission of a closed ring  containing two parallel resistors
was calculated exactly. The Landauer formulation with
single-channel conductors was used to relate the transmission
coefficient to the conductance and the resistors were modelled by
elastic scatterers described by $2 \times 2$ scattering matrices.
Thus, the geometry of the system was similar to that of  Fig. 1a,
except that two general scatterers are placed each on one of the
arms. The transmission amplitudes from the left and from the right
respectively are: $t_i$ and $t'_i,  i = 1,2$, and $r_i,(r'_i),  i
= 1,2$ are the reflection amplitudes on the left (right) of the
scatterer. Notice that time-reversal and current
conservation requirements, which imply $t_i=t'_i$ and
$-t_i/t'^*_i =r_i/r'^*_i$ (the asterisk denotes complex
conjugation), are also satisfied when the geometrical phases of
each path are absorbed in $t_i$, etc. When an AB magnetic flux
$\Phi$ is applied through the center of the ring,  the
transmission and reflection amplitudes pick up the phases
$t_{_1}\rightarrow t_{_1}e^{i\phi}$, $t'_{_1}\rightarrow
t_{_1}e^{-i\phi}$. The three-terminal junctions at the two
connections of the ring to the wires were described by a $3
\times 3$ unitary (chosen real) scattering matrix \cite{Shapiro},
\begin{eqnarray} S=\left(
\begin{array}{ccc}0&-1/\sqrt 2&-1/\sqrt 2 \\
-1/\sqrt 2 &1/2 & -1/2 \\
-1/\sqrt 2 &-1/2&1/2 \end{array} \right ),
\end{eqnarray}
where the
diagonal elements, $S_{ii}$ ($i=1,2,3$) denote the reflection
amplitude of the $i^{th}$ channel, and the off-diagonal elements
$S_{ij}(i\ne j)$ are the transmission amplitudes from channel $j$
to channel $i$. Channel 1 of the left-hand side junction is chosen to be
that of the incoming amplitude (unity) whereas channel 1 of the
right hand side junction is that of the outgoing amplitude ($F$).
For this splitter, no reflection occurs in channel 1 and there is
symmetry between channels 2 and 3. The results  are not expected
to depend qualitatively on the choice of the junction's scattering
matrix, except for the trivial effect that these junctions are
themselves scatterers and add to the total resistance of the
device.
\par From the wave equation for the scattering solution for this model,
it was found in Ref.\cite{GIA} that the total transmission
amplitude, $F$, of the ring is given by
\begin{equation}
F=2{t_{_1}t_{_2}(t'_{_1}+t'_{_2})+t_{_1}(r_{_2}-1)(1-r'_{_2})+t_{_2}(r_{_1}-1)
(1-r'_{_1})\over
(t_{_1}+t_{_2})(t'_{_1}+t'_{_2})-(2-r_{_1}-r_{_2})(2-r'_{_1}-r'_{_2})}.
\label{5.47} \end{equation}

The scatterers ($i = 1,2$) on the two arms are arbitrary elastic ones.
We chose Breit-Wigner forms for them:
\begin{eqnarray} t_i=t'_i =
\frac{i\Gamma_i }{i\Gamma_i  - (E -E_i)}, \\
r_i=r'_i =  \frac{E -E_i}{i \Gamma_i - (E -E_i)},.
\end{eqnarray}

\begin{figure}
\leavevmode \epsfclipon \epsfxsize=8truecm
\vbox{\epsfbox{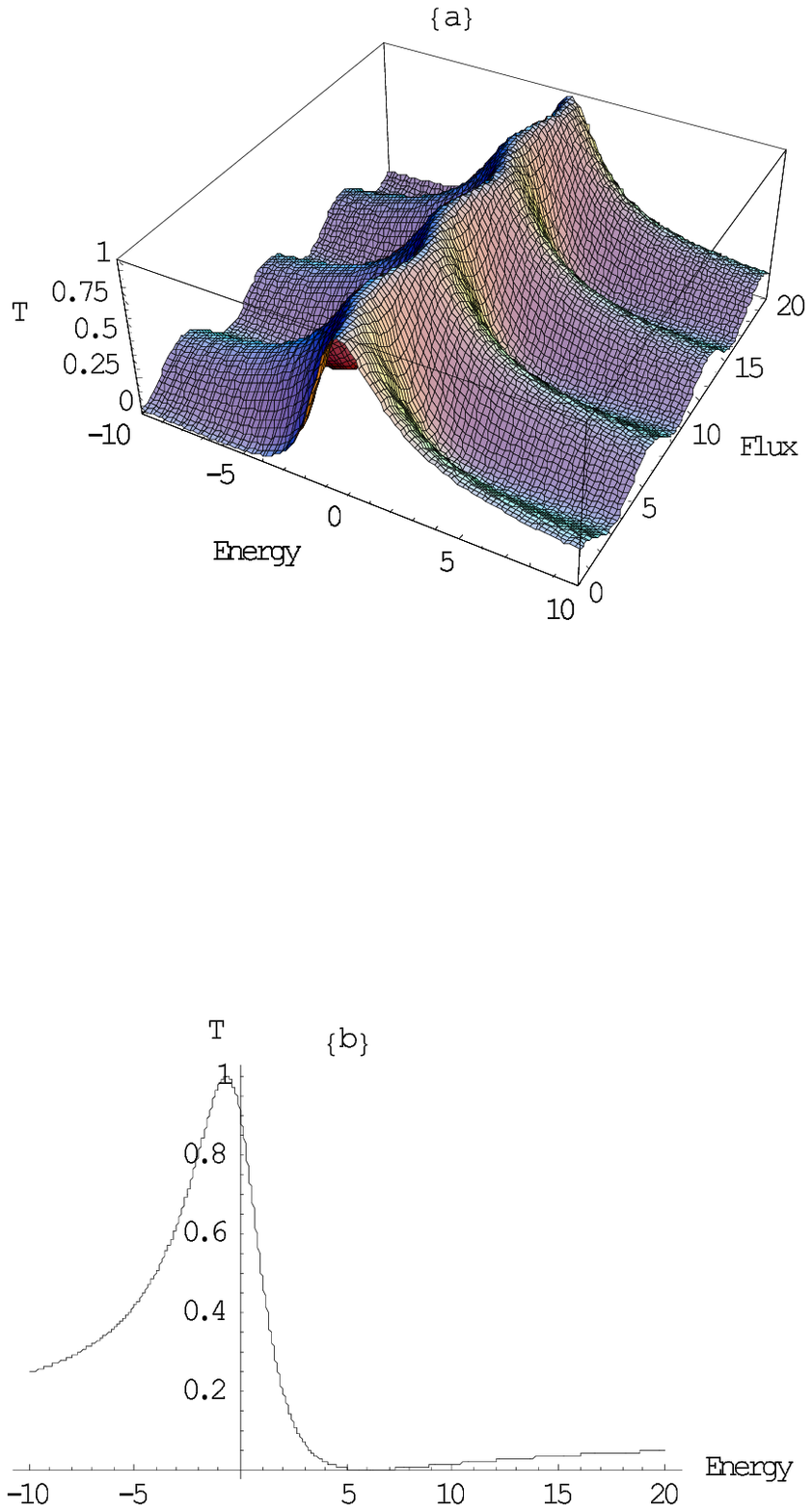}} \caption{{\bf (a)}  Transmission as
function of  $E - E_1$ or ``gate voltage" and $\phi$ for the
closed interferometer of Ref. [4]. Parameters were such that
$\Gamma _1 = 4, E_1 = 0, \Gamma_2 = 2, E - E_2 = 3$ (i.e.  an
energy-independent transmission on the nonresonant branch). Note
the Fano zero at Zero flux and the phase change of $\pi$ across
the resonance. That change happens via the vanishing of the first
Fourier component at a point very close to the resonance. {\bf
(b)} Transmission as function of $E - E_1$ for $\phi = \pi$,
demonstrating a Fano zero. Note its position with respect to the
one in Fig. 3a. }\label{DEQD}
\end{figure}

For scatterer no. 1 we took a resonance at $E_1 = 0$. The
resonance of scatterer 2 was taken further away, so that it can be
regarded as approximately non-resonant around the resonance of 1.
Results for $T = |F|^2$ as  function of energy and flux are depicted
in Fig. \ref{DEQD}a. The $h/e$-periodic  (period of $2 \pi$ in
$\phi$) AB oscillation is clearly seen. This oscillation is even
in $\phi$, as it should. We display only the $\phi \ge 0$ part, in
order to show clearly the Fano-type zero at $\phi = 0$. In
addition, one may also observe how the first harmonic (period $2
\pi$) of the oscillation vanishes near the resonance, i.e. the
sign of that Fourier component reverses there. This simply means
that (except for higher harmonics) the behavior around $\phi = 0$
changes from a maximum to a minimum of $|F|^2$. This is how the
phase change of $\pi$ around a resonance manifests itself for this
interferometer which satisfies unitarity and the Onsager
relationships. For completeness, we display in  Fig. \ref{DEQD}b
the transmission as function of energy at a flux of $\pi$, which
also shows the Fano zero, as expected. This zero occurs for a
positive energy, while the same model (see Fig. 3a later) produces
a Fano zero for a negative energy at vanishing
flux. This opposite
behavior for the two special values of the flux
is consistent with ref. \cite{Fano}.

\subsection{The open two-path interferometer with a DEQD on one of its arms}

As indicated in the introduction,  we now embed the DEQD  as our
effective scattering element on one of the arms of a large {\it
open} AB interferometer (see Fig. 1b) assumed to be designed so
that its transmission is obtained from just adding two  paths. We
have now simply added  the transmission amplitude, $t_{DEQD}$  as
obtained from the above results, to that of a far-from-resonance
and energy-independent scatterer on the other arm. (The latter
models a conducting path which has an energy-independent
transmission amplitude given by $t_{par}$). Now, we plotted the
dimensionless conductance of the whole interferometer as a
function of energy and flux in Fig. 3b.

\begin{figure}
\leavevmode \epsfclipon \epsfxsize=8truecm
\vbox{\epsfbox{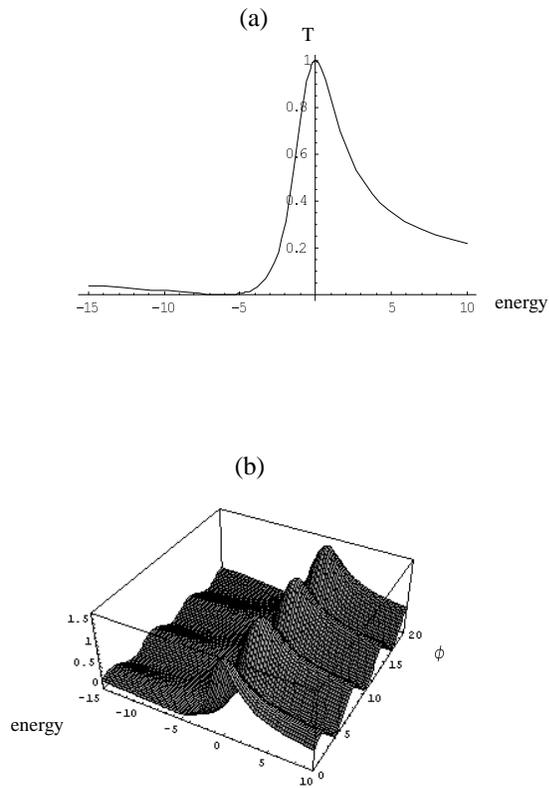}} \caption{{\bf (a)} Transmission as a
function of ``gate voltage"  for a DEQD consisting of a QD with a
single level at $0$ with width of $4$, in dimensionless units, in
parallel with a nonresonant branch with $t = 2i /(2i - 3); r =
3/(2i - 3)$, (the same as that used in fig. 2a). Note the Fano-type zero at about $-5$. {\bf (b)}
Transmission as a function of ``gate voltage" {\it and flux} for
the same DEQD embedded in a two-path interferometer with a
transmitter of $t_{par} = .3$, mimicking experimental data on a
two-path AB interferometer. Note that the phase of the AB
oscillation is changing around the peak but it is coming back and
it is the same far on the two sides of the peak. This is due to
the Fano-type zero and may offer an explanation for the phase
lapse of Ref. [16]. }
\end{figure}
We point out  that {\it
unlike what happened in the closed ``miniature" interferometer
constituting the DEQD in Fig. 2a}, the phase of the AB oscillation
is changing around the peak but it is ``coming back" and becomes
the same far on its two sides. Therefore, the phase increase of
$\pi$ around the resonance is ``rewound" by the zero in the
transmission of the DEQD. This rewinding happens sharply as a
function of energy, when the  flux-dependence vanishes and changes
sign. Thus, the Fano zero in the DEQD produces very interesting
effects in the open AB interferometer.

\section{Simple tight-binding models} \label{Model}

\subsection{A triangular interferometer}
In this section we calculate the scattering matrix through a
triangle, containing an Aharonov-Bohm flux, see Fig. 1c \cite{Entin-Wohlman}. The leads
are described by a tight-binding model, with hopping amplitudes
$-J$, and zero on-site energies, such that the energy of the
scattered electron is
\begin{eqnarray}
\epsilon_{q}=-2J\cos qa.
\end{eqnarray}
The leads are connected to the triangle at its sites $A$ (left)
and $B$ (right). The third site $D$ is the ``quantum dot" (QD).
The direct hopping amplitude
between $A$ and $B$ is denoted $-V$, and is real. The hopping
amplitude from $A$ to $D$ is denoted $-V_{A}e^{i\phi_{A}}\equiv
-\bar{V}_{A}$, and the hopping amplitude from $D$ to $B$ is
denoted $-V_{B}e^{i\phi_{B}}\equiv -\bar{V}_{B}$.
 The total Aharonov-Bohm flux is
\begin{equation}
\Phi=\frac{e}{\hbar c}(\phi_{A}+\phi_{B})=\frac{e}{\hbar c}\phi.
\end{equation}
The Hamiltonian of the system is
\begin{eqnarray}
{\cal H}={\cal H}_{0}+{\cal H}_{1},
\end{eqnarray}
where ${\cal H}_{0}$ is the Hamiltonian of the ordered chain (with
$V$ replaced by $J$), and the isolated QD (with site energy $\epsilon_0$), and
\begin{eqnarray}
{\cal H}_{1}(\ell ,m)&=&-\bar{V}_{A}\delta_{\ell
A}\delta_{mD}-(\bar{V}_{A})^{\ast}\delta_{\ell
D}\delta_{mA}\nonumber\\
&-&\bar{V}_{B}\delta_{\ell
D}\delta_{mB}-(\bar{V}_{B})^{\ast}\delta_{\ell
B}\delta_{mD}\nonumber\\
&-&(V-J)(\delta_{\ell A}\delta_{mB}+\delta_{\ell B}\delta_{mA}).
\label{pert}
\end{eqnarray}
The scattering solution is denoted $\Psi (n)$. It is given by
\begin{eqnarray}
\Psi (n)=\Psi_{0}(n)+\sum_{\ell m}G(n,\ell ){\cal H}_{1}(\ell
,m)\Psi_{0}(m),
\end{eqnarray}
where $\Psi_{0}$ is the incoming wave, so that
\begin{eqnarray}
\Psi_{0}(D)&=&0,\nonumber \\ \noindent\Psi_{0}(n)&\simeq& e^{\pm
iqna},
\end{eqnarray}
the ($\pm$ represents a wave incident from the left or from the
right).

The Green function is calculated from the equation
\begin{eqnarray}
G(n,n')=G_{0}(n,n')+\sum_{\ell m}G(n,\ell ){\cal H}_{1}(\ell
,m)G_{0}(m,n'),
\end{eqnarray}
where $G_{0}$ is the Green function of ${\cal H}_{0}$, so that
\begin{eqnarray}
G_{0}(D,A)&=&G_{0}(A,D)=G_{0}(D,B)=G_{0}(B,D)=0,\nonumber\\
G_{0}(A,A)&=&G_{0}(B,B)\equiv g=\frac{1}{2iJ\sin qa},\nonumber\\
G_{0}(A,B)&=&G_{0}(B,A)=ge^{iqa}.
\end{eqnarray}
The matrix elements of the Green function are calculated in
the appendix. We next express the scattering matrix in terms of
$G(D,D)$.

\subsection{Calculation of the scattering matrix}

For the sake of concreteness,  we assume that the origin is at the
mid-point between the sites $A$ and $B$. Thus
\begin{eqnarray}
\Psi_{0}=e^{\pm iqa(n-1/2)},
\end{eqnarray}
so that for the wave incident from the left
\begin{eqnarray}
\Psi_{0}(A)=e^{-iqa/2},\ \ \Psi_{0}(B)=e^{iqa/2},\ \ \Psi
(A)=e^{-iqa/2}+re^{iqa/2},\ \ \Psi (B)=te^{iqa/2},
\end{eqnarray}
and for the wave incident from the right
\begin{eqnarray}
\Psi_{0}(A)=e^{iqa/2},\ \ \Psi_{0}(B)=e^{-iqa/2},\ \ \Psi
(A)=t'e^{iqa/2},\ \ \Psi (B)=e^{-iqa/2}+r'e^{iqa/2}.
\end{eqnarray}
>From the general equations for the scattered wave,
\begin{eqnarray}
\Psi (A)&=&\Psi_{0}(A)-(\bar{V}_{A})^{\ast}G(A,D)
\Psi_{0}(A)-\bar{V}_{B}G(A,D)\Psi_{0}(B)\nonumber\\
&-&(V-J)G(A,A)\Psi_{0}(B)-(V-J)G(A,B)\Psi_{0}(A),\nonumber\\
\Psi (B)&=&\Psi_{0}(B)-(\bar{V}_{A})^{\ast}G(B,D)
\Psi_{0}(A)-\bar{V}_{B}G(B,D)\Psi_{0}(B)\nonumber\\
&-&(V-J)G(B,A)\Psi_{0}(B)-(V-J)G(B,B)\Psi_{0}(A).
\end{eqnarray}
>From all the above, we obtain
\begin{eqnarray}
1+re^{iqa}&=&2iJ\sin
qa\frac{G(D,D)}{\frac{V^{2}}{J}e^{iqa}-Je^{-iqa}}\times\Bigl
(\epsilon_{q}-\epsilon_{0}+\frac{V_{B}^{2}}{J}e^{iqa}\Bigr
),\nonumber\\
1+e^{iqa}r'&=&2iJ\sin
qa\frac{G(D,D)}{\frac{V^{2}}{J}e^{iqa}-Je^{-iqa}}\times\Bigl
(\epsilon_{q}-\epsilon_{0}+\frac{V_{A}^{2}}{J}e^{iqa}\Bigr
),\nonumber\\
t&=&2iJ\sin
qa\frac{G(D,D)}{\frac{V^{2}}{J}e^{iqa}-Je^{-iqa}}\times\Bigl
(\frac{V}{J}(\epsilon_{q}-\epsilon_{0})-\frac{(\bar{V}_{A}\bar{V}_{B})^{\ast}}{J}\Bigr
),\nonumber\\
t'&=&2iJ\sin
qa\frac{G(D,D)}{\frac{V^{2}}{J}e^{iqa}-Je^{-iqa}}\times\Bigl
(\frac{V}{J}(\epsilon_{q}-\epsilon_{0})-\frac{\bar{V}_{A}\bar{V}_{B}}{J}\Bigr
).\label{amplitudes}
\label{scatt}
\end{eqnarray}


It is interesting to note that the information on the QD enters the scattering matrix
via the Green
function on the dot, $G(D,D)$. As shown by Ng and Lee \cite{Ng}, this remains true also
when one includes electron-electron interactions on the QD.
The important fact is that the interference part, reflected by the terms in
brackets in $t$ and $t'$ in Eq.
(\ref{scatt}), has real coefficients except for the
AB phase factor $e^{i\phi}$.

We are now ready to calculate  the transmission. From Eq. ({\ref{GDD}), we write
\begin{eqnarray}
G^{-1}(D,D)&=&\epsilon_{q}-\epsilon_{0}+
\frac{\frac{V_{A}^{2}+V_{B}^{2}}{J}e^{iqa}+2\frac{V}{J}\frac{V_{A}V_{B}}{J}\cos\phi
e^{2iqa}}{1-\Bigl (\frac{V}{J}\Bigr )^{2}e^{2iqa}}\nonumber\\
&&\equiv {\cal A}+i{\cal B},
\label{GDD1}
\end{eqnarray}
with
\begin{eqnarray}
{\cal A}&=&\epsilon_{q}-\epsilon_{0}\nonumber\\
&+&\frac{\frac{V_{A}^{2}+V_{B}^{2}}{J}\cos qa (1-\Bigl
(\frac{V}{J}\Bigr )^{2}) +2\frac{V}{J}\frac{V_{A}V_{B}}{J}\cos\phi
(\cos 2qa -\Bigl (\frac{V}{J}\Bigr )^{2})}{1+\Bigl
(\frac{V}{J}\Bigr )^{4}-2\Bigl (\frac{V}{J}\Bigr )^{2}\cos
2qa},\nonumber\\
{\cal B}&=& \frac{\frac{V_{A}^{2}+V_{B}^{2}}{J}\sin qa (1+\Bigl
(\frac{V}{J}\Bigr )^{2}) +2\frac{V}{J}\frac{V_{A}V_{B}}{J}\cos\phi
\sin 2qa }{1+\Bigl (\frac{V}{J}\Bigr )^{4}-2\Bigl
(\frac{V}{J}\Bigr )^{2}\cos 2qa}.\label{AB}
\end{eqnarray}
We then define the ``Friedel phase" of the QD \cite{Langreth,Ng},
$\phi_{D}$, such that
\begin{eqnarray}
\cot\phi_{D}=-\frac{{\cal A}}{{\cal B}},\ \
(\sin^{2}\phi_{D}=\frac{{\cal B}^{2}}{{\cal A}^{2}+{\cal B}^{2}},\
\ \cos^{2}\phi_{D}=\frac{{\cal A}^{2}}{{\cal A}^{2}+{\cal
B}^{2}}),\label{friedel}
\end{eqnarray}
which gives
\begin{equation}
G(D,D)=-\frac{\sin\phi_{D}}{{\cal B}}e^{i\phi_{D}}.
\end{equation}
The phase $\phi_D$ is usually refered to as the ``intrinsic transmission phase" of the dot.
When the dot is placed on path 1 of the AB interferometer, $\phi_D$ is equal to the
phase $\alpha_1$ mentioned in our introduction.
The transmission of the structure can now be written in the form
\begin{eqnarray}
T&=|t|^2=&\frac{4\sin^{2}qa}{1+\Bigl (\frac{V}{J}\Bigr )^{4}-2\Bigl
(\frac{V}{J}\Bigr )^{2}\cos
2qa}\times\frac{\sin^{2}\phi_{D}}{{\cal B}^{2}}\nonumber\\
&&\times \Biggl\{\Bigl (\frac{V}{J}\Bigr
)^{2}(\epsilon_{q}-\epsilon_{0})^{2}+\Bigl
(\frac{V_{A}V_{B}}{J}\Bigr
)^{2}-2\frac{V}{J}\frac{V_{A}V_{B}}{J}(\epsilon_{q}-\epsilon_{0})\cos\phi\Biggr
\}.
\end{eqnarray}

Let us now manipulate the term in the curly brackets. From the
equations for ${\cal A}$ and ${\cal B}$ we can write
\begin{eqnarray}
\epsilon_{q}-\epsilon_{0}&=&{\cal A}-\frac{1-\Bigl
(\frac{V}{J}\Bigr )^{2}}{1+\Bigl (\frac{V}{J}\Bigr )^{2}}\cot
qa{\cal B}+2\frac{\frac{V}{J}\frac{V_{A}V_{B}}{J}\cos\phi}{1+\Bigl
(\frac{V}{J}\Bigr )^{2}},\nonumber\\
\frac{V_{A}^{2}+V_{B}^{2}}{J}&=& \frac{1+\Bigl (\frac{V}{J}\Bigr
)^{4}-2\Bigl (\frac{V}{J}\Bigr )^{2}\cos 2qa}{1+\Bigl
(\frac{V}{J}\Bigr )^{2}} \frac{{\cal B}}{\sin qa}
-\frac{4\frac{V}{J}}{1+\Bigl (\frac{V}{J}\Bigr
)^{2}}\frac{V_{A}V_{B}}{J}\cos\phi\cos qa .
\end{eqnarray}
Hence the term in the curly brackets becomes
\begin{eqnarray}
\Biggl \{\Biggr \}&=&\Bigl (\frac{V_{A}V_{B}}{J}\Bigr
)^{2}\sin^{2}\phi +\Bigl
[\frac{V}{J}(\epsilon_{q}-\epsilon_{0})-\frac{V_{A}V_{B}}{J}\cos\phi\Bigr
]^{2}\nonumber\\
&=&\Bigl (\frac{V_{A}V_{B}}{J}\Bigr )^{2}\sin^{2}\phi+\Biggl
[\frac{V}{J}{\cal A}-\frac{1-\Bigl (\frac{V}{J}\Bigr
)^{2}}{1+\Bigl (\frac{V}{J}\Bigr )^{2}}\Biggl (\frac{V}{J}{\cal
B}\cot qa+\frac{V_{A}V_{B}}{J}\cos \phi\Biggr )\Biggr ]^{2}.
\end{eqnarray}

Putting all this together, we obtain
\begin{eqnarray}
T&=&\frac{4\sin^{2}qa}{1+\Bigl (\frac{V}{J}\Bigr )^{4}-2\Bigl
(\frac{V}{J}\Bigr )^{2}\cos 2qa}\Biggl \{\Biggl
(\frac{V_{A}V_{B}\sin\phi_{D}\sin\phi}{J{\cal B}}\Biggr
)^{2}\nonumber\\
&+&\Biggl (\frac{V}{J}\Bigl (\cos\phi_{D}+\frac{1-\Bigl
(\frac{V}{J}\Bigr )^{2}}{1+\Bigl (\frac{V}{J}\Bigr )^{2}}\cot
qa\sin\phi_{D}\Bigr )+\frac{1-\Bigl (\frac{V}{J}\Bigr
)^{2}}{1+\Bigl (\frac{V}{J}\Bigr )^{2}}\frac{V_{A}V_{B}}{J{\cal
B}}\cos \phi\sin\phi_{D}\Biggr )^{2}\Biggr \}.
\end{eqnarray}

\subsection{Fano effect}

>From Eq. (\ref{amplitudes}), the transmission amplitude contains
three factors:
\begin{equation}
t=C G(D,D)[V(\epsilon_q-\epsilon_0)-V_AV_Be^{-i\phi}],
\label{tt}
\end{equation}
where $C=2iJ\sin qa/(V^2e^{iqa}-J^2e^{-iqa})$ does not depend on $\phi$ and does not exhibit
any interesting features near the resonance.
The interference is mainly
reflected by the square brackets. Since all the coefficients inside
these brackets are real, the absolute value of these brackets depends only on $\cos\phi$,
as required by Onsager. $\cos\phi$ also appears in $G(D,D)$, see Eq. (\ref{GDD1}).
Expanding $|G(D,D)|^2$ in a Fourier series in $\phi$, one has
\begin{equation}
|G(D,D)|^2= g_0+g_1 \cos\phi+g_2\cos 2\phi+ ...
\end{equation}
Thus, the coefficient of $\cos\phi$ in $T=|t|^2$
is proportional to
\begin{equation}
[V^2(\epsilon_q-\epsilon_0)^2+V_A^2V_B^2]g_1-2VV_AV_B(\epsilon_q-\epsilon_0)g_0.
\end{equation}
For small $V_AV_B$ and/or $\frac{g_1}{g_0}$,
this vanishes at some energy $\epsilon_q$ close to the original
resonance $\epsilon_0$. As explained above, such vanishing can be interpreted as
a ``jump" of the phase shift $\beta$ by $\pi$.

For a flux $\phi=n\pi$ with integer $n$,
the square brackets in Eq. (\ref{tt}) are real, and
the whole square brackets vanish at $(\epsilon_q-\epsilon_0)=(-1)^n
V_AV_B/V$.
This is exactly the Fano effect and it arises from the exact solution,
without
the lowest order golden rule approximation. For small $V_AV_B$ this vanishing of $T$ (and thus of the
conductance) is also close to the original resonance. If $\phi$ is close to
such integer values,
there will be a dip in $G$ as function of $\epsilon_q$. We expect similar dips in $G$
and ``phase slips" by $\pi$ near every resonance.

\subsection{Special Cases}

We now consider the following special cases:

\begin{enumerate}
\item
$V=0$. This is the Ng-Lee model\cite{Ng}.
We find
\begin{eqnarray}
{\cal B}=\frac{V_{A}^{2}+V_{B}^{2}}{J}\sin qa,
\end{eqnarray}
and hence the transmission is
\begin{eqnarray}
T=\sin^{2}\phi_{D}\Bigl
(\frac{2V_{A}V_{B}}{V_{A}^{2}+V_{B}^{2}}\Bigr )^{2},
\end{eqnarray}
which becomes just $\sin^{2}\phi_{D}$ when $V_{A}=V_{B}$.
\item
$V=J$. In this case
\begin{eqnarray}
{\cal
B}=\frac{\frac{V_{A}^{2}+V_{B}^{2}}{J}+2\frac{V_{A}V_{B}}{J}\cos\phi\cos
qa}{2\sin qa},
\end{eqnarray}
and the transmission is
\begin{eqnarray}
T=\cos^{2}\phi_{D}+\Bigl (\frac{V_{A}V_{B}\sin\phi}{J{\cal
B}}\Bigr )^{2}\sin^{2}\phi_{D}.\label{Kang}
\end{eqnarray}
We note that the model of Kang {\it et al.} \cite{Kang} corresponds
to putting either $V_{A}$ or $V_{B}$ equal to zero. Then the
transmission is just $\cos^{2}\phi_{D}$, and the transmission
phase shift at resonance becomes $\pi$. The result (\ref{Kang})
can be put in the form
\begin{eqnarray}
T=\cos^{2}\phi_{D}+\sin^{2}\phi_{D}\Biggl
(\frac{\frac{2V_{A}V_{B}}{J}(\cos^{2}\frac{\phi
-qa}{2}-\cos^{2}\frac{\phi +qa}{2})}{\Bigl
(\frac{V_{A}-V_{B}}{J}\Bigr )^{2}+2\frac{V_{A}V_{B}}{J}\Bigl
(\cos^{2}\frac{\phi +qa}{2}+\cos^{2}\frac{\phi -qa}{2})}\Biggr
)^{2}.
\end{eqnarray}
The maximal transmission is achieved for $V_{A}=V_{B}$, and is a
combination of  $\cos\phi_{D}$ and $\sin\phi_{D}$ terms. Note
especially the way the magnetic field phase, $\phi$, appears here.
The full expression is of course even in $\phi$.
\item
$V\gg J$. In this case
\begin{eqnarray}
{\cal B}\simeq\Bigl (\frac{J}{V}\Bigr
)^{2}\frac{V_{A}^{2}+V_{B}^{2}}{J}\sin qa,
\end{eqnarray}
and to leading order
\begin{eqnarray}
T\rightarrow \sin^{2}\phi_{D}\Bigl
(\frac{2V_{A}V_{B}}{V_{A}^{2}+V_{B}^{2}}\Bigr )^{2},
\end{eqnarray}
namely, the Ng-Lee result. It is very interesting that the limit
of a strong coupling in the tight-binding model is similar to  a
vanishing coupling.
\item
The symmetric case in which $V_{A}=V_{B}=V$. This
assumption does not simplify considerably the expression for the
transmission. However, it is interesting to note that $V$
will drop out from the result for $T$ (except that it enters back
in the definition of $\phi_{D}$).
\end{enumerate}

\subsection{A QD with two resonances}

One way to decorate the dot is to allow more than one state on it. The effects of two
paths in the DEQD is thus replaced by having two levels on the bare dot, $\epsilon_0$
and $\epsilon_1$.
The couplings of the new level to the ``forks" A and B are then modeled by the additional
tight-binding terms,
\begin{eqnarray}
\Delta {\cal H}_1=-V_{1}e^{i\phi_{1}}\delta_{\ell A}\delta_{m D} -V_2e^{i\phi_2}\delta_{\ell
D}\delta_{ mB}+{\rm h.c.},
\end{eqnarray}
where we assume the same total flux, $\phi=\phi_1+\phi_2$.
The paths going through this level simulate the small parallel
conducting branch in the DEQD.

It is straightforward to repeat the calculation of the scattering
matrix. The result is
\begin{eqnarray}
t'&=&2iJ\sin qa e^{iqa}\Biggl [-V+e^{i\phi}\Bigl
(\frac{V_{A}V_{B}}
{\epsilon_{q}-\epsilon_{0}}+\frac{V_{1}V_2}{\epsilon_{q}-\epsilon_{1}}\Bigl
)\Biggr ]{\cal D}^{-1},\nonumber\\
1+r'e^{iqa}&=&-2iJ\sin qa e^{iqa}\Biggl [1+e^{iqa}\Bigl
(\frac{V_{A}^{2}}{\epsilon_{q}-\epsilon_{0}}+
\frac{V_{1}^{2}}{\epsilon_{q}-\epsilon_{1}}\Bigr )\Biggr
]{\cal D}^{-1}, \label{tra}
\end{eqnarray}
with analogous results for $t$ and $r$.
Here, ${\cal D}$ is related to the $2 \times 2$ Green matrix on the two states
of the QD. It depends on $\phi$ only via $\cos\phi$.

It is interesting to note that the interference part, contained in the square brackets
in $t'$, contains the factor
\begin{eqnarray}
{\cal X}=\frac{V_{A}V_B}{\epsilon_{q}-\epsilon_{0}}+
\frac{V_{1}V_2}{\epsilon_{q}-\epsilon_{1}},
\end{eqnarray}
which plays the role of
\begin{eqnarray}
\frac{V_{A}V_B}{\epsilon_{q}-\epsilon_{0}}
\end{eqnarray}
in the previous results.
The Fano effect, i. e. the vanishing of $t'$,  will now occur when
\begin{equation}
{\cal X}(\epsilon_q)=V.
\end{equation}
If $V_AV_B/(V_1V_2)$ were positive, then this equation always has two solutions, one
being between the original resonance and the other being below or above both. However,
usually the two resonances on the QD correspond
to states of opposite parity. In that case, the ratio is negative, and the equation may
have no real solutions, or both solutions between the original resonances (or on one side
of both), or one solution
on each side of both resonances.
This argument can easily be extended to more resonances: ${\cal X}$ simply becomes a sum
of terms like $V_{L,j}V_{R,j}/(\epsilon_q-\epsilon_j)$, and the Fano zeroes either appear
between every pair of original resonances, or are paired alternately between them.
In any case, in most of the cases one will have a Fano zero associated with each original
resonance.

In addition to the Fano point, each resonance is usually also associated with a nearby point
at which the phase $\beta$ jumps by $\pi$, This happens when $\epsilon_q$ obeys the equation
\begin{equation}
(V^2+{\cal X}^2)d_1-2V{\cal X}d_0=0,
\end{equation}
where we used the Fourier expansion $|{\cal D}|^{-2}=d_0 +d_2\cos\phi+...$.

%
%

\section{The generality of the model and its physical consequences}
\label{Physics}

In Sec. \ref{Fano} we stated that the Fano zeroes are rather
universal for AB interferometers. In addition to our
considerations there, we found these zeroes in both the
scattering model of Ref. \onlinecite{GIA}, treated in Sec. III,
and in the tight-binding model of Ref. \onlinecite{Entin-Wohlman}, generalized in
Sec. \ref{Model}.
They also appear for two coupled resonances, many
coupled resonances and with generic parallel conducting branches.
These results are not peculiar to
specific features, such as the ``fork" $3 \times 3$ matrix used in Ref.
\cite{GIA}. We point out  that such zeroes actually
appeared also in Ref. \cite{BIA} which used a different class of
``fork" matrices. More interestingly, the conditions for the effect
are much more general than having a QD on one of the branches. We
find that Fano zeroes appear also by analyzing the tight-binding
model of Ref. \cite{Claudio} as well as in the model of
Ref.\cite{KG}, with ``QD's" on {\it both} branches. Fano-type
zeroes appear in the problem of two levels coupled to reservoirs
\cite{Fano} and such a zero also shows up in the level statistics
(see the inset to Fig. 1 of Ref. \cite{KGS}). Thus, it is unlikely
that the Fano effect will not play a role in the physics of AB
interferometers and coupled QD's.

Fano physics may well be relevant for the understanding of a
number of experimental anomalies found in interferometers
containing quantum dots. The results in section \ref{Model} show
how the transmission phase at resonance (which should also be
valid for a Kondo resonance) changes from $\pi/2$ in the simple
QD to  $\pi$ for a DEQD with $V=J$ (see also Refs. \cite{Kang,Bulka,Hofs}). This
phase shift will be measured in an open AB interferometer which obeys the two-slit
conditions.
A number of specific mechanisms using variations of the Fano model
have in fact also been advanced
\cite{Hist} as to the relevance of the Fano effect to the ``phase
lapse" discovered experimentally
\cite{Yacoby3,Eyal,Schuster,Ji1,Ji2} in AB interferometers
containing QD's. Related particular mechanisms are based on the
superposition of many resonances \cite{Resos}. (This in fact can
be taken as a particular realization of the parallel conducting
channel model). In the simplest model of  two adjacent
Breit-Wigner resonances, one finds that if signs are right, an
approximate zero in the transmission must occur in-between them.
However, within the  Breit-Wigner approximation to the resonances,
the zero is off the real energy axis by about the width of these
resonances (taken, for simplicity to have comparable widths). What
the Fano considerations tell us, however, is that the zero is
exactly on the real axis and therefore the ``phase lapse'' should
be sharp. The Fano-type considerations therefore augment these
models in a significant fashion. Here we tried to treat all these
Fano-related considerations in a simple and unified fashion.
More work is needed, however, to
examine whether the Fano effect  may really offer a valid
explanation for the experimentally observed phase lapses.

\appendix
\section{ calculation of the Green function}

Our aim here is to calculate the $(D,D)$ matrix element of the
Green function. We have
\begin{eqnarray}
G(D,D)&=&G_{0}(D,D)-\bar{V}_{A}G(D,A)G_{0}(D,D)
-(\bar{V}_{B})^{\ast}G(D,B)G_{0}(D,D),\nonumber\\
G(D,A)&=&-(\bar{V}_{A})^{\ast}G(D,D)G_{0}(A,A)-\bar{V}_{B}G(D,D)
G_{0}(B,A)\nonumber\\
&-&(V-J)\Bigl [ G(D,A)G_{0}(B,A)+G(D,B)G_{0}(A,A)\Bigr
],\nonumber\\
G(D,B)&=&-(\bar{V}_{A})^{\ast}G(D,D)G_{0}(A,B)-\bar{V}_{B}G(D,D)
G_{0}(B,B)\nonumber\\
&-&(V-J)\Bigl [ G(D,A)G_{0}(B,B)+G(D,B)G_{0}(A,B)\Bigr ].
\end{eqnarray}
Using the results
\begin{eqnarray}
G_{0}(A,A)&=&G_{0}(B,B)\equiv g=\frac{1}{2iJ\sin qa},\nonumber\\
G_{0}(A,B)&=&G_{0}(B,A)=ge^{iqa},
 \end{eqnarray}
(where it is assumed that $A$ and $B$ are separated by the same
lattice constant as on the leads), we find
\begin{eqnarray}
G(D,D)&&\equiv \Bigl [\epsilon_{q}-\epsilon_{0}-\Sigma (D,D)\Bigr
]^{-1}\nonumber\\
\Sigma
(D,D)&=&\frac{V_{A}^{2}+V_{B}^{2}+2\frac{V}{J}V_{A}V_{B}e^{iqa}\cos\phi}{
\frac{V^{2}}{J}e^{iqa}-Je^{-iqa}},
\label{GDD}
\end{eqnarray}
and
\begin{eqnarray}
G(D,A)=-\frac{G(D,D)}{\frac{V^{2}}{J}e^{iqa}-Je^{-iqa}}\Bigl (
(\bar{V}_{A})^{\ast}+\frac{V}{J}\bar{V}_{B}e^{iqa}\Bigl
),\nonumber\\
G(D,B)=-\frac{G(D,D)}{\frac{V^{2}}{J}e^{iqa}-Je^{-iqa}}\Bigl (
\bar{V}_{B}+\frac{V}{J}(\bar{V}_{A})^{\ast}e^{iqa}\Bigl ).
\end{eqnarray}

The scattering solution requires the matrix elements
\begin{eqnarray}
G(A,D),\ \ G(A,A),\ \ G(A,B),\ \ G(B,D),\ \ G(B,A),\ \ G(B,B).
\end{eqnarray}
We have the equations
\begin{eqnarray}
G(A,D)&=&-\bar{V}_{A}G(A,A)G_{0}(D,D)-(\bar{V}_{B})^{\ast}G(A,B)G_{0}(D,D),
\nonumber\\
G(B,D)&=&-\bar{V}_{A}G(B,A)G_{0}(D,D)-(\bar{V}_{B})^{\ast}G(B,B)G_{0}(D,D),\nonumber\\
G(A,A)&=&g-(\bar{V}_{A})^{\ast}G(A,D)g-\bar{V}_{B}G(A,D)ge^{iqa}-(V-J)g\Bigl
(G(A,A)e^{iqa}+G(A,B)\Bigr ),\nonumber\\
G(A,B)&=&ge^{iqa}-(\bar{V}_{A})^{\ast}G(A,D)ge^{iqa}-\bar{V}_{B}G(A,D)g-(V-J)g\Bigl
(G(A,A)+G(A,B)e^{iqa}\Bigr ),\nonumber\\
G(B,A)&=&ge^{iqa}-(\bar{V}_{A})^{\ast}G(B,D)g-\bar{V}_{B}G(B,D)ge^{iqa}-(V-J)g\Bigl
(G(B,A)e^{iqa}+G(B,B)\Bigr ),\nonumber\\
G(B,B)&=&g-(\bar{V}_{A})^{\ast}G(B,D)ge^{iqa}-\bar{V}_{B}G(B,D)g-(V-J)g\Bigl
(G(B,A)+G(B,B)e^{iqa}\Bigr ).
\end{eqnarray}
The third and fourth equations here give
\begin{eqnarray}
G(A,A)\frac{V}{J}e^{iqa}-G(A,B)&=&-G(A,D)\frac{\bar{V}_{B}}{J}e^{iqa},\nonumber\\
G(A,B)\frac{V}{J}e^{iqa}-G(A,A)&=&
\frac{1}{J}e^{iqa}-\frac{(\bar{V}_{A})^{\ast}}{J}G(A,D)e^{iqa}.
\end{eqnarray}
Likewise, the last two equations yield
\begin{eqnarray}
G(B,B)\frac{V}{J}e^{iqa}-G(B,A)&=&-G(B,D)\frac{(\bar{V}_{A})^{\ast}}{J}e^{iqa},\nonumber\\
G(B,A)\frac{V}{J}e^{iqa}-G(B,B)&=&
\frac{1}{J}e^{iqa}-\frac{\bar{V}_{B}}{J}G(B,D)e^{iqa}.
\end{eqnarray}
Hence, the solutions are
\begin{eqnarray}
G(A,D)&=&-\frac{G(D,D)}{\frac{V^{2}}{J}e^{iqa}-Je^{-iqa}}\Bigl
(\bar{V}_{A}+\frac{V}{J}(\bar{V}_{B})^{\ast}e^{iqa}\Bigr
),\nonumber\\
G(B,D)&=&-\frac{G(D,D)}{\frac{V^{2}}{J}e^{iqa}-Je^{-iqa}}\Bigl
((\bar{V}_{B})^{\ast}+\frac{V}{J}\bar{V}_{A}e^{iqa}\Bigr
),\nonumber\\
G(A,A)&=&\frac{G(D,D)}{\frac{V^{2}}{J}e^{iqa}-Je^{-iqa}}\Bigl
(\epsilon_{q}-\epsilon_{0}+\frac{V_{B}^{2}}{J}e^{iqa}\Bigr
),\nonumber\\
G(B,B)&=&\frac{G(D,D)}{\frac{V^{2}}{J}e^{iqa}-Je^{-iqa}}\Bigl
(\epsilon_{q}-\epsilon_{0}+\frac{V_{A}^{2}}{J}e^{iqa}\Bigr
),\nonumber\\
G(B,A)&=&\frac{G(D,D)}{\frac{V^{2}}{J}e^{iqa}-Je^{-iqa}} \Bigl
(\frac{V}{J}e^{iqa}(\epsilon_{q}-\epsilon_{0})
-\frac{(\bar{V}_{A}\bar{V}_{B})^{\ast}}{J}e^{iqa}\Bigr
),\nonumber\\
G(A,B)&=&\frac{G(D,D)}{\frac{V^{2}}{J}e^{iqa}-Je^{-iqa}} \Bigl
(\frac{V}{J}e^{iqa}(\epsilon_{q}-\epsilon_{0})
-\frac{\bar{V}_{A}\bar{V}_{B}}{J}e^{iqa}\Bigr ).
\end{eqnarray}

\vspace{1.5cm}

{\bf Acknowledgements}: We thank  M. Heiblum, P. W\"olfle, A.
Schiller and D. Sprinzak for helpful conversations. O.E-W,
A.A. and Y.I. thank the Institute for Theoretical Physics at the
University of California, Santa Barbara for its hospitality
when this research was concluded and the paper written. This project
was supported by by the German-Israeli Foundation (GIF), by a
Center of Excellence of the Israel Science Foundation, by the
Albert Einstein Minerva Center for Theoretical Physics at the
Weizmann Institute of Science and by the National Science Foundation
under Grant No.PHY99-07949.

\newpage


\end{document}